\documentstyle[12pt]{article}
\begin{document}
\sloppy
\sloppy
\sloppy

\begin{flushright}{UT-804\\ December, 1997}
\end{flushright}
\vskip 0.5 truecm

\vskip 0.5 truecm

\begin{center}
{\large{\bf  Fluctuation-dissipation theorem and quantum tunneling
with dissipation }}
\end{center}
\vskip .5 truecm
\centerline{\bf Kazuo Fujikawa}
\vskip .4 truecm
\centerline {\it Department of Physics,University of Tokyo}
\centerline {\it Bunkyo-ku,Tokyo 113,Japan}
\vskip 0.5 truecm

\begin{abstract}
We suggest to take the fluctuation-dissipation theorem of Callen and Welton as 
a basis to study quantum dissipative phenomena ( such as macroscopic quantum 
tunneling ) in a manner analogous to the Nambu-Goldstone theorem for spontaneous symmetry breakdown. It is shown that the essential physical contents  of the Caldeira-Leggett model such as the suppression of quantum coherence by Ohmic dissipation are derived from general principles only, namely, the fluctuation-dissipation  theorem and  unitarity  and 
causality (i.e., dispersion relations), without 
referring to an explicit form of the Lagrangian.  An interesting connection 
between quantum tunneling with Ohmic dissipation and the Anderson's orthogonality 
theorem  is also noted. 
\par
\end{abstract}
 
\section{Fluctuation-dissipation theorem}
The quantum dissipative phenomena[1] are related to the study of irreversible processes in quantum statistical mechanics, and as such they have been studied by many authors in the past, in particular, in the framework of linear response theory[2][3]. A combination of dissipation with quantum tunneling adds further interesting aspects to the subject and potentially covers a wide class of phenomena in various fields of physics.  In the analysis of macroscopic quantum tunneling with dissipation, the model defined by Caldeira and Leggett[4] is widely used. 
One of the essential ingredients of their model is to simulate the effects of dissipation by an infinite number of harmonic oscillators. It has been discussed in a convincing way  by Feynman and Vernon [3] and also by Caldeira and Leggett themselves [4] that this 
method  of using an infinite number of harmonic oscillators is in fact quite general. 
Nevertheless, it may be interesting to demonstrate explicitly that the quantum 
dissipative phenomena can be studied in a model independent manner without 
introducing an infinite number of harmonic oscillators. Of course, if one wants to do
without oscillators, one needs an alternative input to define the problem. We 
want to take an  input which is as general as possible and yet  manageable. 

We here  propose  to take the fluctuation-dissipation theorem of Callen and Welton [1] as a basic principle to study quantum dissipative phenomena. It is then 
shown that the essential  physical contents of the Caldeira-Leggett 
model such as the suppression of quantum coherence by Ohmic dissipation are  reproduced by a combination of the fluctuation-dissipation theorem and
dispersion relations ( or unitarity and causality).
By this way  we can  set  the quantum tunneling with dissipation in a broader perspective , and the Caldeira-Leggett model is  recognized  as a minimal Lagrangian  model which incorporates  the essential aspects of macroscopic quantum tunneling with dissipation in the lowest order of perturbation. An interesting 
similarity between quantum tunneling with Ohmic dissipation and the impurity atom 
hopping problem in a metal , which is related to the Anderson's orthogonality 
theorem , is also revealed in the present approach.  

We first recapitulate  the basic reasoning for  the fluctuation-dissipation  theorem [1] in a way which is convenient for our application.  We start with the 
Hamiltonian
\begin{equation}
H = H_{0}(q) + qQ
\end{equation}
where $q$ stands for the freedom we are interested in, which is described by 
$H_{0}(q)$ , and $Q$ describes the freedom associated with the medium (or dissipative system). Note that $Q$ stands for quite complicated dynamical degrees of  freedom. 
The choice of (1) may be justified in the spirit of linear 
response theory [2][3]. 

To specify the dynamics of $Q$ ( and thus the Hamiltonian $H_{0}(Q)$ indirectly), we consider the following gedanken experiment:  we consider the external (given) sinusoidal motion of the variable $q$
\begin{equation}
q_{ext}(t) = q_{ext}(0)\sin \omega t
\end{equation}
with {\em small} $q_{ext}(0)$, and we adopt the interaction Hamiltonian
\begin{equation}
H_{I}^{\prime} = q_{ext}(t)Q
\end{equation}
with the dynamics  $H_{0}(q)$ switched-off. 
The lowest order perturbation  (Fermi's golden rule) with $H_{I}^{\prime}$ gives the transition probability (starting at the state $|E_{n}\rangle$ of the 
$Q$-system )
\begin{eqnarray}
w &=& \frac{\pi}{2 \hbar}q_{ext}(0)^{2}\{ |\langle E_{n} + \hbar \omega |Q| E_{n}\rangle |^{2}\rho (E_{n} + \hbar \omega) \nonumber\\
  && + |\langle E_{n} - \hbar \omega |Q| E_{n}\rangle |^{2}\rho (E_{n} - \hbar \omega)\}
\end{eqnarray}
where $\rho (E_{n})$ stands for the state density of $Q$-system at $E=E_{n}$. The energy absorption rate by the system $Q$ is obtained by multiplying $\hbar \omega$ to the difference of the two terms in (4) , absorption and emission terms, as 
\begin{eqnarray}
 &&\frac{\pi\omega }{2 }q_{ext}(0)^{2}\{ |\langle E_{n} + \hbar \omega |Q| E_{n}\rangle |^{2}\rho (E_{n} + \hbar \omega) \nonumber\\
  &&\ \ \ \ \  - |\langle E_{n} - \hbar \omega |Q| E_{n}\rangle |^{2}\rho (E_{n} - \hbar \omega)\}
\end{eqnarray}
The energy absorption per unit time at temperature $T$ is thus given by 
\begin{eqnarray}
&& \frac{\pi\omega }{2 }q_{ext}(0)^{2}\sum_{n}\{ |\langle E_{n} + \hbar \omega |Q| E_{n}\rangle |^{2}\rho (E_{n} + \hbar \omega) \nonumber\\
  &&\ \ \ \ \  - |\langle E_{n} - \hbar \omega |Q| E_{n}\rangle |^{2}\rho (E_{n} - \hbar \omega)\}
f(E_{n})
\end{eqnarray}
with $f(E_{n})$ the (normalized ) Boltzmann factor , which satisfies $f(E_{n} + \hbar\omega )/
f(E_{n}) = \exp (-\hbar\omega / kT )$.  This final expression (6) for the dissipative power can also be written as 
\begin{eqnarray}
Power &=&  \frac{\pi\omega }{2 }q_{ext}(0)^{2}\omega \int_{0}^{\infty} dE \rho (E) f(E)\{ |\langle E + \hbar \omega |Q| E\rangle |^{2}\rho (E + \hbar \omega) \nonumber\\
  && - |\langle E - \hbar \omega |Q| E\rangle |^{2}\rho (E - \hbar \omega)\}
\nonumber\\
&=&  \frac{\pi\omega }{2 }q_{ext}(0)^{2}\omega (1-e^{-\hbar\omega /kT})\nonumber\\
&&\times \int_{0}^{\infty} dE \{ |\langle E + \hbar \omega |Q| E\rangle |^{2}\rho (E + \hbar \omega)\rho (E) f(E)  \}   
\end{eqnarray} 
where we used $|\langle E_{n} - \hbar \omega |Q| E_{n}\rangle| = |\langle E_{n} |Q| E_{n} -  \hbar \omega \rangle |$ and $f(E + \hbar \omega )/f(E) = e ^{-\hbar\omega/kT}$ after a shift in integration variable.

On the other hand, the presence of the {\em induced} dissipation (7) for the 
sinusoidal given  motion $q_{ext}(t)$ suggests a phenomenological dissipative force (or reaction) acting on $q_{ext}(t)$
\begin{equation}
F = - R(\omega) \dot{q}_{ext}(t)
\end{equation}
where $R(\omega)$ stands for the dissipative coefficient (resistance) for the 
sinusoidal motion with frequency $\omega$. We do not ask the microscopic 
dynamics which produces the specific form of the dissipative force in (8).  The energy dissipation by this 
induced friction is given by 
\begin{eqnarray}
Energy \ \ dissipation& /& unit\ \  time \nonumber\\
&=& - \overline{F \dot{q}_{ext}(t)}\nonumber\\
&=& R(\omega)\overline{\dot{q}_{ext}(t)^{2}} = \frac{\omega^{2}}{2}
R(\omega)q_{ext}(0)^{2}
\end{eqnarray}
where the bar shows the time averaging. 
 
 If one compares the macroscopic expression (9) with the microscopic expression (7) ,  one finds 
 \begin{equation}
 R(\omega) = \frac{\pi}{\omega}(1-e^{-\hbar\omega/kT})
\int_{0}^{\infty}|\langle E+\hbar\omega |Q|E\rangle |^{2}\rho (E+ \hbar\omega )\rho (E)f(E)dE
\end{equation}
Therefore one obtains 
\begin{eqnarray}
&&(\frac{2}{\pi})\frac{\hbar\omega}{2}\coth (\hbar\omega/2kT)
R(\omega)  \nonumber\\                 
&=& \hbar\int_{0}^{\infty}(1 + e^{-\hbar\omega/kT})|\langle E+\hbar\omega |Q|E\rangle |^{2}\rho (E+ \hbar\omega )\rho (E)f(E)dE\nonumber\\
&=& \hbar\int_{0}^{\infty}\{ |\langle E+\hbar\omega |Q|E\rangle |^{2}\rho (E+ \hbar\omega ) \nonumber\\
&&+ |\langle E - \hbar\omega |Q|E\rangle |^{2}\rho (E - \hbar\omega )\}\rho (E)
f(E)dE
\end{eqnarray}
where we used $|\langle E_{n} + \hbar \omega |Q| E_{n}\rangle| = |\langle E_{n} |Q| E_{n} +  \hbar \omega \rangle |$, $\rho (E) = 0 $ for $E<0$,  and $f(E + \hbar \omega )/f(E) = e ^{-\hbar\omega/kT}$ combined with a shift in integration variable.
This is a local version of the fluctuation-dissipation theorem for each value of $\omega$. 

If one integrates the above relation (11) with respect to $\omega$, one has
\begin{eqnarray}
&&\frac{2}{\pi}\int_{0}^{\infty}\frac{\hbar\omega}{2}\coth (\hbar\omega/2kT)
R(\omega)d\omega  \nonumber\\                 
&=& \int_{0}^{\infty}\{ \int_{0}^{\infty} |\langle E+\hbar\omega |Q|E\rangle |^{2}\rho (E+ \hbar\omega )d(\hbar\omega) \nonumber\\
&&+ \int_{0}^{\infty}|\langle E - \hbar\omega |Q|E\rangle |^{2}\rho (E - \hbar\omega )d(\hbar\omega)\}\rho (E)f(E)dE\nonumber\\
&=&  \int_{0}^{\infty}\langle E |Q^{2}|E\rangle \rho (E)f(E)dE  \nonumber\\
&\equiv& \langle Q^{2}\rangle 
\end{eqnarray}
after a change of the order of integration in $E$ and $\hbar\omega$.
We thus obtain the well-known theorem of Callen and Welton [1]
\begin{eqnarray}
\frac{2}{\pi}\int_{0}^{\infty}\frac{\hbar\omega}{2}\coth (\hbar\omega/2kT)
R(\omega)d\omega &=&\frac{2}{\pi}\int_{0}^{\infty}E(\omega, T)
R(\omega)d\omega \nonumber\\
&=& \langle Q^{2}\rangle 
\end{eqnarray}
The factor 
\begin{equation}
E(\omega, T) = \frac{1}{2}\hbar\omega + \frac{\hbar\omega}{e^{\hbar\omega/(kT)}
-1}
\end{equation}
stands for the mean energy of the {\em bosonic} harmonic oscillator with the zero point 
energy $ \frac{1}{2}\hbar\omega $  included. The coefficient $R(\omega)$ may generally depend on the temperature, but in the above analysis  $R(\omega)$ is 
assumed to be effectively independent of the temperature in the region we are interested in. In the context of the present analysis,  one should  rather regard eq.(10) as a specification of the medium for a given $R(\omega)$.

Although there are some criticisms [5] about the basis for the linear response theory [2] and the fluctuation-dissipation theorem itself [1], we take the relation (13)  as the starting point of our specification of the medium which induces the effective frictional force in (8), instead of specifying the 
Hamiltonian $H_{0}(Q)$ explicitly. One may  draw the following intuitive picture for the theorem (13):

The existence of dissipation suggests the presence of some (collective) excitation in the medium. The spectrum of such excitation (or force field) is reflected 
in the resistance $R(\omega)$, which measures the {\em on-shell radiation} of such 
an effective excitation mode( see eqs.(8) and (9)). Namely, the excitation exists for the frequency $\omega$ for which $R(\omega) \neq 0$. Also, such effective  excitation, if classically 
recognizable, should necessarily be {\em bosonic}. Combined with a dimensional  analysis, it is thus natural to have a general expression (13).

In this note , we propose to understand the relation (13) in a manner analogous to  the 
Nambu-Goldstone theorem [6] which asserts the inevitable presence of a 
massless excitation if a continuous symmetry is spontaneously broken.
In contrast, the relation (13) asserts that the presence of dissipation specified
by $R(\omega)$ inevitably leads to the effective  excitation whose energy 
spectrum  at temperature $T$ is specified by $(2/\pi)E(\omega, T)R(\omega)$,
at least in the region $T\simeq 0$.

In the following, we demonstrate  that some of the essential physical contents of the 
Caldeira-Leggett model[4] are derived from general principles only, nemely, 
the theorem (13) and unitarity and  
causality (i.e., dispersion relations ) without referring to an explicit form 
of Lagrangian. The present analysis is partly motivated by a field theoretical re-formulation [7] of the Caldeira-Leggett model, where the unitarity and causality 
are manifestly exhibited. In fact, the actual analysis of dissipative tunneling below  closely follows that in Ref.[7]. 

\section{Quantum tunneling with dissipation}
 
We are mainly interested in the quantum tunneling phenomena ( rather than  thermally assisted tunneling), and thus we work on the case of zero temperature. The fluctuation-dissipation theorem in the form  (13) is still  useful to provide a definite regularization of the manipulations in (7) $\sim$ (11),  even though we work at  zero temperature.

We start with the calculation of the decay rate of the state $|n\rangle$ with 
energy $E_{n}$ to the state $|m\rangle$ with energy $E_{m}$ for the q-system
described by $H_{0}(q)$ in (1). By taking $H_{I} = qQ$ as a perturbation, the lowest order perturbation formula gives
\begin{equation}
w(n\rightarrow m + \hbar\omega) = \frac{2\pi}{\hbar}|\langle m|q|n\rangle|^{2}
|\langle \hbar\omega |Q|0\rangle|^{2}\delta (E_{n}- E_{m} - \hbar\omega)
\end{equation}
or if one takes a sum over final states of the environmental system Q, we obtain
\begin{eqnarray}
&& w(n\rightarrow m + \hbar\omega)\nonumber\\
&& = \frac{2\pi}{\hbar}|\langle m|q|n\rangle|^{2}\int |\langle \hbar\omega |Q|0\rangle|^{2}\delta (E_{n}- E_{m} - \hbar\omega)\rho (\hbar\omega)d(\hbar\omega)\nonumber\\
&&= \frac{2\pi}{\hbar}|\langle m|q|n\rangle|^{2}\frac{2}{\pi}\frac{\hbar\omega}{2}R(\omega)\frac{1}{\hbar}|_{\hbar\omega = E_{n} - E_{m}}
\end{eqnarray}
where we used the local version of the fluctuation-dissipation theorem (11) in  the form  
\begin{equation}
|\langle \hbar\omega |Q|0\rangle|^{2}\rho (\hbar\omega)d(\hbar\omega) = 
\frac{2}{\pi}\frac{\hbar\omega}{2}R(\omega)d\omega
\end{equation}
for $\omega \geq 0$ at $T = 0$.  We thus obtain the decay rate 
\begin{equation}
 w(n\rightarrow m + \hbar\omega) = \frac{2}{\hbar}|\langle m|q|n\rangle|^{2}
(\frac{E_{n}-E_{m}}{\hbar})R(\frac{E_{n}-E_{m}}{\hbar})\theta (E_{n}-E_{m}) 
\end{equation}
with the step function $\theta (x)$. 

In the following we first specialize in the Ohmic dissipation, for which we have $R(\omega) = \eta = constant$, and 
\begin{equation}
 w(n\rightarrow m + \hbar\omega) = \frac{2\eta}{\hbar}|\langle m|q|n\rangle|^{2}
(\frac{E_{n}-E_{m}}{\hbar})\theta (E_{n}-E_{m}) 
\end{equation}
In particular, if the q-system is described by a simple harmonic oscillator, 
$H_{0}(q) = \frac{1}{2M}p^{2} + \frac{M\omega^{2}}{2}q^{2}$, we obtain 
\begin{eqnarray}
w(n\rightarrow m + \hbar\omega) &=& \frac{2\eta}{\hbar}(\frac{\hbar}{2M\omega})
(\frac{E_{n}-E_{m}}{\hbar})|_{\hbar\omega = E_{n}-E_{m}}\nonumber\\
&=& \frac{\eta}{M}\delta_{m,n-1}
\end{eqnarray}
The half-decay width is obtained from (20) as  
\begin{eqnarray}
\frac{1}{2}\Gamma_{n}& =&\frac{1}{2}\hbar \sum_{m} w(n\rightarrow m + \hbar\omega)\nonumber\\
&=&\frac{1}{2}\hbar \frac{\eta}{M}   
\end{eqnarray}
This  expression of $\Gamma_{n}$ is consistent with the equation of motion for the damped oscillator
\begin{equation}
M\ddot{q} + \eta\dot{q} + M\omega^{2}q = 0
\end{equation}
and also with our starting assumption in (8). 

If one denotes the proper self-energy correction to the state $|n\rangle$ of the q-system with $H_{0}(q)|n\rangle = E_{n}|n\rangle$ by $\Sigma_{n}(E)$, i.e., if the Green's function is written as $\langle n|i/(E - \hat{H}+ i\epsilon )|n\rangle =
i[ E - E_{n} +i\epsilon + \Sigma_{n}(E) ]^{-1}$, one can write  a dispersion relation 
\begin{equation}
\Sigma_{n}(E) = \frac{1}{\pi}\int_{0}^{\Lambda}\frac{Im \Sigma_{n}(E^{\prime})dE^{\prime}}{E^{\prime} - E -i\epsilon}
\end{equation}
with an infinitesimal positive constant $\epsilon$ and $\Lambda$ is a cut-off
parameter. The total Hamiltonian is written as $\hat{H} \equiv H_{0}(q) + qQ + H_{0}(Q)$. The imaginary (or absorptive) part of $\Sigma_{n}(E)$ is given by (16) and (19) as ( in the lowest order in $\eta$)
\begin{eqnarray}
Im \Sigma_{n}(E)&\equiv& \frac{1}{2}\Gamma_{n}(E)\nonumber\\
&=& \pi\sum_{m}|\langle m|q|n\rangle|^{2}\frac{2}{\pi}\frac{\hbar\omega}{2}R(\omega)\frac{1}{\hbar}|_{\hbar\omega = E - E_{m}}\nonumber\\
&=& \frac{\eta}{\hbar}\sum_{m}|\langle m|q|n\rangle|^{2}(E - E_{m})\theta(E -E_{m})
\end{eqnarray}
which is a manifestation of unitarity (  an analogue of optical theorem).
The dispersion  relation ( up to all orders in perturbation of $qQ$) follows from the fact that $\Sigma_{n}(E)$ is analytic in the upper-half plane due to causality, and $\Sigma_{n}(E)$ is real and regular along the negative real axis as a result of positive definiteness of the Hamiltonian $H_{0}(q) + H_{0}(Q)$. One may continue $\Sigma_{n}(E)$  to the lower-half plane along the negative real axis by 
$\Sigma_{n}(E -i\epsilon) = \Sigma_{n}(E + i\epsilon)^{\star}$. The imaginary part of $\Sigma_{n}(E)$ arises since the state $|n\rangle$ is not an  eigenstate of the total $\hat{H}$. (The relations (23) and (24) are of course derived from second order perturbation theory. What we emphasized here is that the relation (23) is of more general validity.)  

We then have 
\begin{eqnarray}
 \Sigma_{n}(E) &=& \frac{\eta}{\pi\hbar}\sum_{m}|\langle m|q|n\rangle|^{2}\int_{0}^{\Lambda}\frac{(E^{\prime}- E_{m}) \theta(E^{\prime} - E_{m})dE^{\prime}}{E^{\prime} - E -i\epsilon}\nonumber\\
&=& \frac{\eta}{\pi\hbar}\sum_{m}|\langle m|q|n\rangle|^{2}\int_{E_{m}}^{\Lambda}\frac{(E^{\prime}- E_{m}) dE^{\prime}}{E^{\prime} - E -i\epsilon}\nonumber\\ 
&\simeq&\frac{\eta}{\pi\hbar}\sum_{m}|\langle m|q|n\rangle|^{2}\int_{0}^{\Lambda}\frac{(\hbar\omega) d(\hbar\omega)}{\hbar\omega +
E_{m} - E -i\epsilon}\nonumber\\ 
&=&\frac{\eta}{\pi\hbar}\sum_{m}|\langle m|q|n\rangle|^{2}\{ \hbar\Lambda \nonumber\\
&& +
(E-E_{m})\int_{0}^{\Lambda}\frac{ d\omega}{\omega + (E_{m} - E)/\hbar -i\epsilon}\} 
\end{eqnarray}
 
If we apply a very specific subtraction procedure to the above dispersion relation by
subtracting the constant $\Lambda$ term in the last expression , we obtain 
\begin{equation}
\Sigma_{n}(E)= \frac{\eta}{\pi\hbar}\sum_{m}|\langle m|q|n\rangle|^{2}(E-E_{m})\int_{0}^{\Lambda}\frac{ d\omega}{\omega + (E_{m} - E)/\hbar -i\epsilon}
\end{equation}
The above subtraction convention of Caldeira and Leggett [4] amounts to setting the 
contribution of the state $|m\rangle$ to $\Sigma_{n}(E)$ to vanish at $E=E_{m}$:
In other words, the dissipative interaction does not influence  $\Sigma_{n}(E)$ at 
the vanishing frequency $\omega = 0$. Physically this means that the shape of the potential for q-system (which is defined at the static limit) is not influenced by the term $\frac{\eta\Lambda}{\pi}\langle n|q^{2}|n\rangle$ induced by dissipation. This prescription of  subtraction  is the crux of the {\em macroscopic} quantum tunneling formulated in [4].

For sufficiently large $\Lambda$, we have 
\begin{eqnarray}
&&\Sigma_{n}(E_{n})\nonumber\\
&&= \frac{\eta}{\pi\hbar}\sum_{m}|\langle m|q|n\rangle|^{2}(E_{n}-E_{m})[\ln |\frac{\Lambda - (\frac{E_{n}-E_{m}}{\hbar})}{(E_{n}-E_{m})/\hbar}| + i\pi\theta (E-E_{m})] \nonumber\\
&&\simeq \frac{\eta}{\pi\hbar}\{ -\frac{\hbar^{2}}{2M}\ln \frac{\Lambda}{\mu}\nonumber\\
&& \ \ \ \ \  + \sum_{m}|\langle m|q|n\rangle|^{2}(E_{n}-E_{m})[ \ln |\frac{\hbar\mu}{E_{n}-E_{m}}| + i\pi\theta (E_{n}-E_{m}) ]\}
\end{eqnarray}
where we used the sum rule 
\begin{equation}
\sum_{m}\langle n|q|m\rangle \langle m|q|n\rangle (E_{m} - E_{n}) = \frac{\hbar^{2}}{2M}
\end{equation}
which follows from 
\begin{equation}
[q, [q, H_{0}(q)]] = - \frac{\hbar^{2}}{M}
\end{equation}
independently of the detailed form of the potential $V(q)$ for the q-system, 
$H_{0}(q) = \frac{1}{2M}p^{2} + V(q)$ . 

The corrected energy eigenvalue of the state $|n\rangle$ is thus given by 
\begin{eqnarray}
E_{n}^{(1)} &=& E_{n} - Re \Sigma_{n}(E_{n})\nonumber\\
&=& E_{n} + \frac{\hbar\eta}{2\pi M}\ln \frac{\Lambda}{\mu}\nonumber\\
&& \ \ \  - \frac{\eta}{\pi\hbar}\sum_{m}|\langle m|q|n\rangle|^{2}(E_{n}-E_{m})\ln |\frac{\hbar\mu}{E_{n}-E_{m}}| 
\end{eqnarray}
where $\mu$ specifies an arbitrary renormalization point, 
and the half-width of the state $|n\rangle$ is given by 
\begin{eqnarray}
\frac{\Gamma_{n}}{2} &=& Im \Sigma_{n}(E_{n})\nonumber\\
&=& \frac{\eta}{\hbar}\sum_{m}|\langle m|q|n\rangle|^{2}(E_{n}-E_{m})\theta (E_{n} - E_{m})
\end{eqnarray}
which is in fact the input to the dispersion relation (23). These relations are identical to those in a field theoretical approach to the quantum tunneling with dissipation[7]. 

Our formula (25) is valid for the general potential problem , and it is not 
restricted to  quantum tunneling with dissipation. We now specialize in the 
tunneling problem ( to be precise,  quantum coherence) for which we have (isolated) nearly degenerate
states arising from quantum tunneling . From the formula for $E_{n}^{(1)}$ in ( 30), we obtain [7]
\begin{eqnarray}
\epsilon_{1}&\equiv& E_{1}^{(1)}-E_{0}^{(1)}\nonumber\\
&\simeq& \epsilon - \frac{2\eta}{\pi\hbar}|\langle 0|q|1\rangle|^{2}\epsilon
\ln (\frac{\hbar\omega_{0}}{\epsilon})
\end{eqnarray}
for the lowest two states , which arise from tunneling splitting, of a {\em deep }
double-well potential $V(q)$  with $\epsilon \equiv E_{1} - E_{0}$ ; the frequency $\omega_{0}$, which plays a role of cut-off parameter,  is the curvature at the bottom of  the symmetric double-
well potential.  The energy splitting $\epsilon_{1}$ stands for the fundamental order parameter of macroscopic quantum tunneling.  It is shown that the effects of other states can be neglected for a deep potential. 

The ``renormalization group'' equation
\begin{eqnarray}
\beta &=& \omega_{0}\frac{d\epsilon_{1}(\omega_{0})}{d\omega_{0}}\nonumber\\
&=& - \bar{\eta}\epsilon \simeq - \bar{\eta}\epsilon_{1}(\omega_{0})
\end{eqnarray}
with 
\begin{equation}
\bar{\eta} \equiv \frac{2\eta}{\pi\hbar}|\langle 0|q|1\rangle|^{2} 
\end{equation}
gives an ``improved'' formula 
\begin{equation}
\epsilon_{1} = \epsilon (\frac{\epsilon}{\hbar\omega_{0}})^{\bar{\eta}}
\end{equation}
if one normalizes $\epsilon_{1}(\hbar\omega_{0} = \epsilon) = \epsilon$.
The self-consistency ansatz [8] by setting infrared cut-off at $\epsilon_{1}$, 
namely 
$\epsilon_{1} = \epsilon (\frac{\epsilon_{1}}{\hbar\omega_{0}})^{\bar{\eta}}$ 
then gives rise to the  well-known formula [4]
\begin{equation}
\epsilon_{1} = \epsilon (\frac{\epsilon}{\hbar\omega_{0}})^{\frac{\bar{\eta}}{
1-\bar{\eta}}}
\end{equation}
which suggests a strong  suppression of quantum coherence for $\bar{\eta}\rightarrow 1$.\\
\\
{\bf  Super-Ohmic dissipation }\\

We next analyze the case of super-Ohmic dissipation
\begin{equation}
R(\omega) = \eta\omega^{2}
\end{equation}
which is related to the emission of electromagnetic fields in the vacuum [1]. 
The calculation proceeds in a manner identical to that of the Ohmic dissipation, and we obtain
\begin{equation}
 \Sigma_{n}(E) = \frac{\eta}{\pi\hbar}\sum_{m}|\langle m|q|n\rangle|^{2}\int_{0}^{\Lambda}\frac{(E^{\prime}- E_{m})(\frac{E^{\prime}- E_{m}}{\hbar})^{2} \theta(E^{\prime} - E_{m})dE^{\prime}}{E^{\prime} - E -i\epsilon}
\end{equation}
with
\begin{equation}
Im \Sigma_{n}(E)= 
 \frac{\eta}{\hbar}\sum_{m}|\langle m|q|n\rangle|^{2}(E - E_{m})(\frac{E - E_{m}}{\hbar})^{2}\theta(E - E_{m})
\end{equation}
After the subtraction as in (26) above, we obtain
\begin{eqnarray}
 \Sigma_{n}(E)& =& \frac{\eta}{\pi}\sum_{m}|\langle m|q|n\rangle|^{2}
\{\frac{\Lambda^{2}}{2}(\frac{E - E_{m}}{\hbar}) + \Lambda(\frac{E - E_{m}}{\hbar})^{2}\nonumber\\
&&\ \ +  (\frac{E - E_{m}}{\hbar})^{3}[\ln |\frac{\hbar\Lambda}{E - E_{m}}| +
i\pi\theta (E- E_{m})]\}
\end{eqnarray}
We thus have
\begin{eqnarray}
 &&\Sigma_{n}(E_{n}) = -\frac{\hbar\eta}{4\pi M}\Lambda^{2} + \frac{\eta}{\pi M^{2}}\Lambda \langle n|p^{2}|n\rangle \nonumber\\
&& + \frac{\eta}{\pi}\sum_{m}|\langle m|q|n\rangle|^{2}
 (\frac{E_{n} - E_{m}}{\hbar})^{3}[\ln |\frac{\hbar\Lambda}{E_{n} - E_{m}}| +
i\pi\theta (E_{n}- E_{m})]\}
\end{eqnarray}
where we used the sum rules (28) and 
\begin{equation}
\sum_{m}|\langle m|q|n\rangle|^{2}(E_{n} - E_{m})^{2} = \frac{\hbar^{2}}{M^{2}}
\langle n|p^{2}|n\rangle
\end{equation}
The second term in the right-hand side of equation (41) modifies the kinetic 
term of the Hamiltonian $H_{0}(q) = \frac{1}{2M}p^{2} + V(q)$  for q-system, and it may be natural to subtract it away in the spirit of the macroscopic quantum tunneling of  Caldeira and Leggett [4]. 

We thus finally obtain the corrected eigenvalue to the order linear in $\eta$
\begin{eqnarray}
E_{n}^{(1)} &=& E_{n} - Re \Sigma_{n}(E_{n})\nonumber\\
&=& E_{n}+ \frac{\hbar\eta}{4\pi M}\Lambda^{2} 
 - \frac{\eta}{\pi}\sum_{m}|\langle m|q|n\rangle|^{2}
 (\frac{E_{n} - E_{m}}{\hbar})^{3}\ln |\frac{\hbar\Lambda}{E_{n} - E_{m}}| 
\end{eqnarray}
where the second term proportional to $\Lambda^{2}$ does not influence the observable energy splitting between two energy eigenvalues.

It has been shown in Ref.[7] that one  obtains the energy splitting of lowest two levels in the deep double-well potential $V(q)$ from (43) as (see also eq.(32))
\begin{equation}
\epsilon_{1} \simeq \epsilon + \frac{\eta\omega_{0}}{2\pi M}[\epsilon^{\prime} 
+ 3\epsilon \ln (\frac{\Lambda}{\omega_{0}})]
\end{equation}
where $\epsilon^{\prime}$ is the tunneling energy splitting of the 3rd and 4th
energy levels in the deep double-well potential. If one accepts the above subtraction procedure, which appears to be reasonable in the spirit of the Caldeira-Leggett model, the tunneling splitting ( and consequently, tunneling itself) is rather enhanced by the super-Ohmic dissipation for a {\em small} $\eta$. 

\section{Discussion and Conclusion}
 We have demonstrated that some of the essential  physical contents of the Caldeira-Leggett model for the macroscopic quantum tunneling with dissipation can be reproduced from the fluctuation dissipation theorem of Callen and Welton (13) and the 
dispersion relations ( or unitarity and causality). The existence of dissipation implies a fluctuation in the force field, which in turn modifies the tunneling frequency via dispersion relations. The present approach is  consistent with a field theoretical formulation of the Caldeira-Leggett model in Ref.[7], where the unitarity and causality are explicitly incorporated. An explicit form of the Lagrangian contains  more information and thus one can explicitly evaluate corrections to 
the energy eigenvalue in the second order of  $\eta$, for example [7].   

We note that the formula (30) is also valid for a more general class of interaction between $q$- system $H_{0}(q)$ and $Q$- system $H_{0}(Q)$ defined by 
\begin{equation}
H_{I}= F(q)Q
\end{equation}
In this case the formula (30) is replaced by 
\begin{eqnarray}
E_{n}^{(1)}
&=& E_{n} - \frac{\eta}{\pi\hbar}\sum_{m}|\langle m|F(q)|n\rangle|^{2}(E_{n}-E_{m})\ln |\frac{\hbar\Lambda}{E_{n}-E_{m}}| 
\end{eqnarray}
The dynamics of $H_{0}(Q)$, which is characterized by $R(\omega)$, is specified by a diagnosis by means of $H_{I}^{\prime}= q_{ext}(t)Q$ in (3), but we can choose a more general interaction (45) for an actual analysis. The analysis of 
quantum coherence such as (32) is however dependent on the detailed properties 
of $F(q)$.

If one draws an analogy of quantum dissipative phenomena with  spontaneous symmetry breakdown characterized by the Nambu-Goldstone theorem [6], which specifies the zero mass excitation of the vacuum,  the present approach corresponds to a specification of  
Nambu-Goldstone bosons and their physical implications on the basis of current
algebra and low-energy theorem. On the other hand, the  Caldeira-Leggett model
corresponds to an  effective field theory (or non-linear $\sigma$-model) for 
Nambu-Goldstone bosons; an efffective field theory, once an explicit form of the  Lagrangian is given, naturally contains more information, though {\em not}
 all of the properties of the effective Lagrangian stand for the  generic properties of Nambu-Goldstone bosons. In the context of quantum dissipation, it is known [3]
that any system of dissipative $Q$-freedom can be approximated by an infinite number of harmonic oscillators if the second order in  perturbation (i.e.,  linear response as in (4) )
yields satisfactory accuracy. But of course, the true dynamics of $Q$-system 
could be quite different from an infinite number of harmonic oscillators. As for  practical physical implications of this consideration, the behavior of  (36) for 
$\bar{\eta}\rightarrow 1$ is more dependent on the detailed model of  $Q$-system, since it is sensitive to the detailed dynamics in the  higher orders of   $\bar{\eta}$; our analysis of (36) is valid for  $\bar{\eta} \ll 1$ .  

We also note an interesting similarity between the quantum tunneling with Ohmic dissipation and the impurity atom hopping problem in a metal [9]. In the latter problem, the atom is dressed by conduction electrons which give rise to the effect similar to dissipation. The effective dressed hopping frequency, $\Delta_{eff}/\hbar$, is obtained at zero temperature as [9]
\begin{equation}
\Delta_{eff} = \Delta (\frac{\Delta}{D})^{K}
\end{equation}
where  $\Delta/\hbar$ stands  for the bare hopping frequency and $D$ is the 
Fermi energy with $K$ a positive coupling constant. In fact, the entire derivation of (47) in [9] is almost identical to our calculation of the Ohmic dissipation in (35).  The interaction term in the atom hopping has a structure 
\begin{equation}
H_{I} = \sigma_{x}\sum_{k^{\prime},k} N_{k^{\prime}, k}a_{k^{\prime}}^{\dagger}a_{k}
\end{equation}
in the lowest two-level truncation of the atomic states described by the Pauli
matrix $\sigma_{x}$, and $a_{k}$ and 
$a_{k}^{\dagger}$ stand for annihilation and creation operators of the conduction
electron. The second order perturbation theory gives rise to a relation corresponding to the dispersion relation in (23).
The assumption of slowly varying  $|N_{k^{\prime}, k}|^{2}$ after the angular integral of  $k$ and $k^{\prime}$ gives rise   to an (effective) 
Ohmic dissipation ( i.e., $R(\omega) = constant$ in our notation ), as is expected for a charge movement in a metal. In other 
words, $H_{I}$ in (48) is effectively simulated by  the fluctuation-dissipation theorem of Callen and Welton with 
$\hbar\omega = \epsilon (k^{\prime}) - \epsilon (k)$ in the restricted domain of frequency $\hbar\omega$ and temperature. Kondo [9] suggests that 
the above reduction of the hopping frequency (47) is understood as a variant 
of Anderson's orthogonality theorem [10] ( i.e., the overlap integral of two 
Fermi gas states decreases when a large number of degrees of freedom are involved). This view 
point may provide an alternative interesting physical picture for  the 
suppression of quantum tunneling with Ohmic dissipation [11], although a precise 
correspondence between these two ideas, Ohmic dissipation and orthogonality theorem ,  remains to be clarified.  
  
A theoretical basis for the fluctuation-dissipation theorem compared with the Nambu-Goldstone theorem is less solid,  mainly due to the difficulty to describe a deviation from a thermal equilibrium, but if one observes dissipative phenomena characterized  by a well-defined $R(\omega)$, one may specify the essential physical properties of the dissipative medium by the relation (10) and the theorem (13). A  message of the present note   is that this information alone is sufficient to analyze the essential physical effects of dissipation on macroscopic  quantum coherence.  
\\ 

I thank M. Ueda for illuminating discussions on the fluctuation- dissipation
theorem and for calling Ref.[9] to my attention.

\end{document}